\documentclass[preprint,11pt, authoryear]{elsarticle}
\usepackage{apalike}
\usepackage{cite}
\usepackage{relsize}
\usepackage{tikz}
\usepackage{lipsum,adjustbox}
\usepackage{amssymb,amsfonts,bm,amsthm}
\usepackage{amsmath}
\usepackage{algorithmic}
\usepackage{graphicx}
\usepackage{textcomp}
\usepackage{xcolor}

\newtheorem{thm}{Theorem}

\newtheorem{coro}{Corollary}
\usepackage{subfigure}
\usepackage{tcolorbox}
\usepackage{booktabs}
\usepackage{tabularx}
\usepackage{multicol}
\usepackage{multirow}
\usepackage{longtable}
\usepackage{nomencl}
\usepackage{enumitem,kantlipsum}

\usepackage{geometry}

\newgeometry{twoside}

\makenomenclature

\makeatletter
\newcommand{\pushright}[1]{\ifmeasuring@#1\else\omit\hfill$\displaystyle#1$\fi\ignorespaces}
\newcommand{\pushleft}[1]{\ifmeasuring@#1\else\omit$\displaystyle#1$\hfill\fi\ignorespaces}
\makeatother

\journal{European Journal of Operational Research}

\begin{document}

\newpageafter{author}

\begin{frontmatter}



\title{Optimizing Bidding Curves for Renewable Energy in Two-Settlement Electricity Markets} 

\author[dongwei]{Dongwei Zhao\corref{cor1}}
\cortext[cor1]{Corresponding author: Dongwei Zhao, email: dongwei.zhao@anl.gov}

\affiliation[dongwei]{organization={Argonne National Laboratory},
            city={Lemont},
            state={IL},
            country={USA}}

\author[stefanos]{Stefanos Delikaraoglou}
\affiliation[stefanos]{organization={Laboratory for Information and Decision Systems, Massachusetts Institute of Technology},
            city={Cambridge},
            state={MA},
            country={USA}}

\author[vladimir]{Vladimir Dvorkin}
\affiliation[vladimir]{organization={Department of 
 Electrical Engineering and Computer Science , University of Michigan},
            city={Ann Arbor},
            state={MI},
            country={USA}}

\author[alberto]{Alberto J. Lamadrid L.}
\affiliation[alberto]{organization={Department of Economics and
Department of Industrial and Systems Engineering, Lehigh University},
            city={Bethlehem},
            state={PA},
            country={USA}}

\author[audun]{Audun Botterud}
\affiliation[audun]{organization={Laboratory for Information and Decision Systems, Massachusetts Institute of Technology},
            city={Cambridge},
            state={MA},
            country={USA}}

\clearpage
\newpage

\begin{abstract}
Coordination of day-ahead and real-time electricity
markets is imperative for cost-effective electricity supply and also to provide efficient incentives for the energy transition. Although stochastic market designs feature the least-cost coordination, they are incompatible with current deterministic markets. This paper proposes a new approach for compatible coordination in two-settlement markets based on benchmark bidding curves for variable renewable energy. These curves are optimized based on a bilevel optimization problem, anticipating per-scenario responses of deterministic market-clearing problems and ultimately minimizing the expected cost across day-ahead and real-time markets. Although the general bilevel model is challenging to solve, we theoretically prove that a single-segment bidding curve with a zero bidding price is sufficient to achieve system optimality if the marginal cost of variable renewable energy is zero, thus addressing the computational challenge. In practice, variable renewable energy producers can be allowed to bid multi-segment curves with non-zero prices. We test the bilevel framework for both single- and multiple-segment bidding curves under the assumption of fixed bidding prices. We leverage duality theory and McCormick envelopes to derive the linear programming approximation of the bilevel problem, which scales to practical systems such as a 1576-bus NYISO system. We benchmark the proposed coordination and find absolute dominance over the baseline solution, which assumes that renewables agnostically bid their expected forecasts. We also demonstrate that our proposed scheme provides a good approximation of the least-cost, yet unattainable in practice, stochastic market outcome. 
\end{abstract}


\begin{keyword}

Continuous optimization \sep  bilevel optimization  \sep electricity market \sep renewable energy \sep  bidding curve



\end{keyword}

\end{frontmatter}



\section*{Nomenclature}
\addcontentsline{toc}{section}{Nomenclature}
\subsection*{Sets and Indexes}

\noindent $\Lambda$:  Set of transmission lines, indexed by $(n,m)$, where $n/m$ is sending/receiving end of a transmission line

\noindent $\Omega$: Set of VRE generation scenarios, indexed by $\omega$

\noindent $\mathcal{I}$:   Set of conventional units, indexed by $i$ or $j$

\noindent $\mathcal{I}^\text{FS}$: Subset of $\mathcal{I}$ with fast-startup generation units

\noindent $\mathcal{I}^\text{SL}$: Subset of $\mathcal{I}$ with slow-startup generation units

\noindent $\mathcal{K}$: Set of VRE generation units, indexed by $k$ 

\noindent $\mathcal{N}$: Set of grid buses, indexed by $n$

\noindent $\mathcal{T}$: Set of market operation time slots, indexed by $t$

\noindent $\mathcal{S}$: Set of VRE bidding curve segments, indexed by $s$ 

\noindent $(\cdot)_{n}$: mapping of $(\cdot)$ into the set of buses

\subsection*{Parameters}

\noindent $C_{i}$:   Variable generation cost of conventional units 

\noindent $C_{i}^0$:  No-load cost of conventional units 

\noindent $C_{i}^{\text{SU}}$: Start-up cost of conventional units 

\noindent $C^{\text{sh}}$:  Cost of load shedding (value of lost load) 

\noindent $C_{i}^{\text{U/D}}$:  Real-time  up-/downward re-dispatch costs

\noindent $\overline{F}_{n,m}$: Transmission line capacities

\noindent $L_{n,t}$:  Day-ahead load forecast

\noindent $L_{n,t,\omega}$: Real-time demand realization 

\noindent $\overline{P}_{i}$: Maximum generation capacity of conventional units

\noindent $\underline{P}_{i}$: Minimum generation capacity of conventional units

\noindent  ${R}_{i}^{\text{U/D}}$:  Upward/downward ramping capacity of conventional units

\noindent  $\widetilde{W}_{k,t,\omega}$: Real-time VRE power realization

\noindent  $\overline{W}_{k}$: VRE power capacity  

\noindent $x_{n,m}$: Transmission line reactance

\subsection*{Decision Variables}

\noindent $\delta_{n,t}^{\text{DA}}$: Day-ahead voltage angle

\noindent $c_{i,t}^\text{DA}$: Day-ahead start-up cost of conventional units

\noindent $p_{i,t}^{\text{C}}$: Day-ahead generation of conventional units

\noindent  $p_{k,t,s}^{\text{W}}$: Day-ahead generation of VRE units  

\noindent  $u_{i,t}^\text{DA}$:  Day-ahead commitment status of conventional units

\noindent  $C_{k,t,s}^\text{W}$: Day-ahead bidding cost of VRE units over bidding curve

\noindent  $W_{k,t, s}$:  Day-ahead bidding quantity of VRE units over bidding curve

\noindent  $\delta_{n,t,\omega}^{\text{RT}}$: Real-time voltage angle

\noindent  $c_{i,t,\omega}^\text{RT}$: Real-time start-up cost of conventional units

\noindent  $l_{n,t,\omega}^{\text{sh}}$:  Real-time  shedding of electrical loads 

\noindent  $p_{k,t,\omega}^{\text{W,cr}}$: Real-time VRE power curtailment

\noindent  $r_{i,t,\omega}^{\text{U/D}}$: Real-time up-/downward re-dispatch of conventional units

\noindent  $u_{i,t,\omega}^\text{RT}$: Real-time commitment status of conventional units

\section{Introduction}

\subsection{Background and Motivation}
The transition to zero-carbon power systems is driving the large-scale integration of variable renewable energy sources (VRE) in restructured electricity markets. In the United States, for example, the New York Independent System Operator (NYISO) and the California Independent System Operator (CAISO) are implementing measures to align with state policies mandating that 100\% of electricity come from zero-carbon sources by 2040 \citep{nyisowinplan} and 2045 \citep{caisowinplan}, respectively.

Despite these efforts, current electricity market designs remain rooted in the technical and economic frameworks developed for dispatchable fossil-fueled generators. Short-term electricity markets typically operate in two settlements or stages \citep{fundamentals}: a day-ahead market (DAM), cleared before actual operations, and a real-time market (RTM), which addresses imbalances closer to delivery. Dispatchable fossil-fueled generators are generally capable of meeting DAM commitments with minimal RTM adjustments. However, the increasing penetration of VRE introduces greater risks to electricity supply due to its inherent variability and uncertainty, posing significant challenges for efficiently managing RTM imbalances \citep{zhou2022price}.

Since the DAM scheduling will determine RTM redispatch, efficiently scheduling VRE in the DAM is of vital importance to achieving market efficiency. 
Notably, VRE basically has zero marginal costs with uncertain outputs. If VRE producers bid with the true marginal cost, their bids will be dispatched in priority in the DAM based on the merit order. However, although VRE has zero marginal cost in the DAM, their uncertain output may lead to significant RTM re-dispatch costs. To reflect the RTM re-dispatch cost, some electricity markets allow VRE producers to bid prices higher than zero. For example, VRE producers can submit multi-segment bidding curves to indicate price and quantity preferences in NYISO \citep{nyisowindbid}. These bidding curves are usually private decisions of VRE producers.  However, if these curves are not properly constructed to reflect the true cost in DAM and RTM, they may jeopardize market efficiency.

The above considerations motivate this work to study the key question:   \textit{What are optimal DAM bidding curves for VRE producers in two-settlement electricity markets to achieve the expected least system cost?}

\subsection{Main Contributions}
 To answer the above question, we formulate and analyze a bilevel optimization framework that jointly optimizes bidding prices and quantities of multi-segment bidding curves. We explore the solution method to solve the problem for large-scale systems. Specifically, we make the following four contributions:

\begin{enumerate} 
\item \textit{Bilevel optimization framework for bidding curves:} To the best of our knowledge, this is the first work to optimize the bidding curves for VRE in the two-settlement electricity market, which aims to achieve the minimum system cost. We present a bilevel optimization framework to optimize DAM bidding curves, which anticipates the real-time redispatch cost induced due to VRE uncertainty and minimizes the expected system cost of day-ahead and real-time dispatch.

\item \textit{Optimality guaranteed by single-segment bidding curves:} The bilevel problem that jointly optimizes prices and quantities of bidding curves is challenging to solve. 
However, we prove theoretically that from the system perspective,  a single-segment bidding curve with a zero bidding price and optimal quantity is sufficient to achieve optimality. 

 \item \textit{Simulations on large-scale systems:} VRE producers in practice can be allowed to bid multi-segment curves with non-zero prices. Given bidding prices, we test the bilevel framework for single-segment and multi-segment bidding curves, respectively. We adopt our previously proposed method of strong duality and McCormick envelopes \citep{zhao2022uncertainty} to relax the bilevel optimization problem into a linear program (LP), which can solve at scale, e.g., a  1576-bus NYISO system. 

 \item \textit{Practical insights:}  Although VRE producers can bid with different prices under single- or multi-segment bidding curves in the DAM, we demonstrate that the bilevel framework can effectively adjust the bidding quantities corresponding to the bidding prices and achieve good performance in terms of the system cost. Hence, our proposed bilevel framework can potentially serve as a tool to guide or regulate VRE producers' bidding strategies.
\end{enumerate}

\subsection{Related Work}

To enhance coordination between DAM and RTM operations, extensive literature has proposed various stochastic dispatch methods, integrating stochastic optimization, chance-constrained optimization, and robust optimization into electricity market clearing to address uncertainties. Studies such as \citep{wong2007pricing, zavala2017stochastic, Kazempour2018Market, Exizidis2019market} employ scenario-based stochastic optimization to clear electricity markets. This approach generally ensures minimal expected system costs but relies on precise probability distributions of uncertain parameters. Furthermore, scenario-based stochastic optimization often struggles to guarantee revenue adequacy and cost recovery for individual scenarios \citep{morales2012pricing}. Some studies, including \citep{kuang2018, dvorkin2019chance,mieth2020risk}, have explored market mechanisms based on chance-constrained optimization.  While chance constraints offer a straightforward way to control risk tolerance, their reformulation assumptions—such as the linear dependence of optimization variables on forecast errors—are restrictive, potentially leading to conservative and suboptimal solutions  \citep{10068745}. Other works, such as \citep{ cobos2018network,velloso2019two}, have applied robust optimization to electricity market clearing. This approach does not require explicit probability distributions of VRE but it optimizes solely for the worst case, potentially leading to poor performance under typical operating conditions.

The stochastic dispatch approaches in the above literature cleared the day-ahead and real-time operations in a joint manner. Despite achieving optimal coordination between the day-ahead and real-time stages, this design is not directly compatible with existing market structures, where day-ahead and real-time markets are cleared independently in a deterministic manner. To address this issue, the work \citep{morales2014electricity} proposed an adjustment of DAM wind-energy quantities using bilevel optimization, which approximates the stochastic market solution within the sequential deterministic energy-only market design. Follow-up work in \citep{dvorkin2018setting, delikaraoglou2019optimal, viafora2020dynamic} extended this bilevel approach to reserve and energy co-optimization.  More recently, some literature has focused on informing forecast models on the cost of forecast errors, also leading to enhanced temporal coordination between day-ahead and real-time markets.  The works in \citep{zhang2023deriving, zhang2024toward, dvorkin2024regression} developed machine learning algorithms that yield cost-informed predictions to minimize the cost of real-time redispatch. The work \citep{zhang2023deriving, zhang2024toward} focused on a bilevel optimization problem from the system perspective while the work \citep{dvorkin2024regression} solved a game-theoretic model leveraging VRE producers' strategic behaviors.

Our work also adopts the bilevel framework to coordinate the DAM and RTM to minimize the expected system cost, but it differs from \citep{morales2014electricity, dvorkin2018setting, delikaraoglou2019optimal,viafora2020dynamic,zhang2023deriving, zhang2024toward} in several crucial ways: First, all the above works have both LPs at the upper level and lower level, focusing on the quantity adjustment of generation units. Our bilevel model jointly optimizes bidding quantities and prices, which involves a bilinear objective at the upper level, but we can prove that the optimality can be achieved through a simplified bilevel model under a single quantity segment with a zero bidding price. Second, the works \citep{morales2014electricity,dvorkin2018setting,delikaraoglou2019optimal,viafora2020dynamic} solved bilevel optimization problems by formulating mixed-integer linear programming (MILP) problems based on Karush--Kuhn--Tucker (KKT) conditions. However, these methods were only tested on small-scale systems, such as a 24-bus system \citep{morales2014electricity}. Although \citep{zhang2023deriving, zhang2024toward} adopted a learning method to solve the bilevel problem, it was only tested on a small 4-generator system. In our previous work \citep{zhao2022uncertainty},  we proposed a method based on strong duality \citep{boyd2004convex} and McCormick envelope \citep{mccormick1976computability}, which can relax the bilevel optimization problem into an LP problem and be applied to large-scale systems. However,  our previous work \citep{zhao2022uncertainty} only focused on the single-segment bidding curve under zero bidding prices. In this work, we will apply this approach to the single- and multi-segment curves under different bidding prices. We demonstrate that this more general approach can still achieve good scalability and accuracy results on large-scale systems. e.g., a 1576-bus NYISO system. 

In terms of bidding curve optimization for VRE producers, the literature \citep{dai2017optimum, ghavidel2019risk, zhao2019storage, hu2021price, zhao2023bid} focused on the perspective of suppliers, which aims to optimize revenues or profits based on price-taking or price-making models. To the best of our knowledge, no literature studies the optimal bidding curve of VRE  in the two-settlement electricity market, with a focus on minimizing the system cost. Our work addresses this research gap.

The rest of the paper is organized as follows. We first introduce the market-clearing model in Section \ref{section:market}. Then, we formulate a general bilevel framework for VRE bidding curve optimization in Section \ref{section:bilevel} and show that the optimality is achieved by a simplified bilevel model in Section \ref{section:equiv}. We provide numerical demonstrations for the case when bidding prices are given in Section \ref{section:givenprice}. Finally, Section  \ref{section:conclusion} concludes.

\section{Electricity Market Model} \label{section:market}
This section presents models for day-ahead and real-time market clearing, aligned with the foundational structure of the U.S. electricity market. Regarding modeling assumptions, the network topology is included considering linear DC power flows. Conventional generation units are modeled with linear operating costs. We assume that system demand is inelastic and assigned a high Value of Lost Load (VoLL) to reflect the penalty for unserved demand. To preserve convexity in the market-clearing models, we relax binary unit commitment (UC) decisions, resulting in a simplified UC formulation \citep{kasina2014comparison, kazempour2017value}. Additionally, the model distinguishes between fast-start and slow-start generators but omits minimum on- and off-time constraints for simplicity.

We denote the operation horizon for market clearing by $\mathcal{T}$ and the set of VRE producers by $\mathcal{K}$. To address system uncertainties, we introduce a scenario set $\Omega$, capturing the discrete probability distribution of real-time VRE generation and system demand variations.\footnote{We do not include energy storage in the current model. On one hand, the bidding curves of energy storage are not trivial to set up. On the other hand, the inter-temporal constraints may significantly increase the computation time of the bilevel model. It is interesting and challenging to optimize the bidding curves of energy storage and VRE jointly and we leave it as future work.}
 
Next, we introduce the model of VRE bidding curves and formulate market-clearing optimization models.

\subsection{Bidding Curves of VRE}
 We assume that VRE producers can bid multi-segment curves with non-negative prices to reflect the RTM re-dispatch cost. For conventional generation units, we assume that they just bid the  true cost and capacity in the market.

\begin{figure}[t]
	\centering
         {\includegraphics[width=3in]{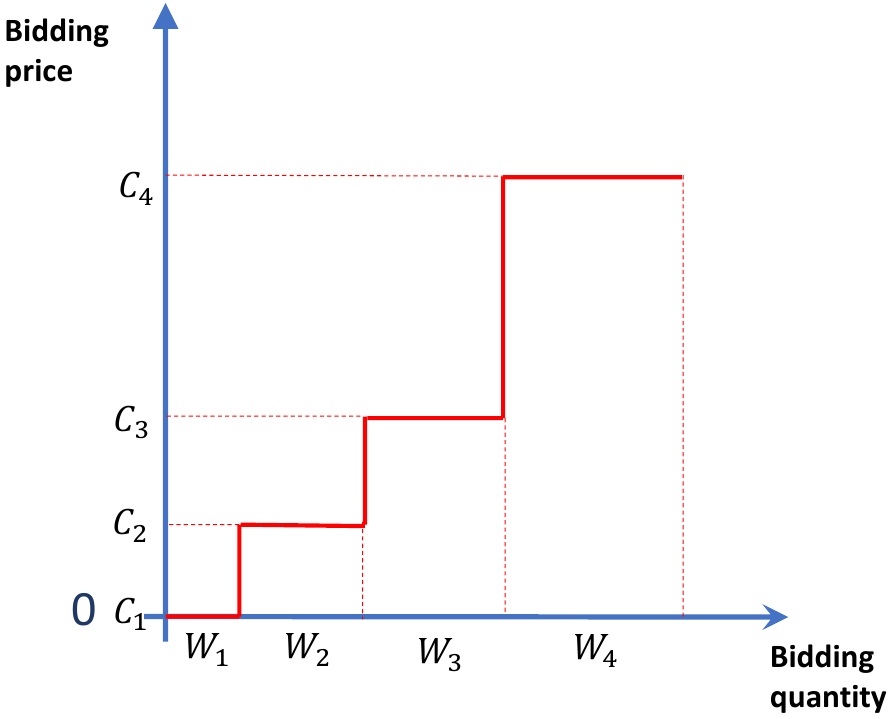}}
         \vspace{-1ex}
	\caption{Bidding curve example for VRE producer. }
	\label{fig:capacity_cruve_w1}
\end{figure}

We model the bidding curve of each VRE producer at each hour and each bus, which consists of multiple segments of prices and quantities.    One example of a bidding curve with 4 segments is shown in Fig. \ref{fig:capacity_cruve_w1}. Suppose an $S$-segment bidding curve with the set of segments denoted by $\mathcal{S}$. For each segment $s\in \mathcal{S}$ of VRE supplier $k \in \mathcal{K}$ at hour $t\in \mathcal{T}$, we denote the bidding price\footnote{This work only considers the nonnegative prices. In this work, we empirically set the bidding price for each segment, e.g., by approximating the aggregate supply cost curves of conventional units in the current market.} by $C_{k,t,s}^{\text{W}}$ and the bidding quantity by $W_{k,t,s}$. Without loss of generality, we let $0\leq C_{k,t,1}^{\text{W}}\leqslant C_{k,t,2}^{\text{W}},...,\leqslant C_{k,t,S}^{\text{W}}$ $\leqslant \bar{C}^{\text{W}}$, where $\bar{C}^{\text{W}}$ is an upper bound that is sufficiently large.  We let $(\bm{C}^{\text{W}},\bm{W})=$ $(C_{k,t,s}^{\text{W}},W_{k,t,s},\forall s\in \mathcal{S}, t\in \mathcal{T},k\in \mathcal{K})$. We assume that the number of segments $S$ is given as a parameter.  


\subsection{Day-Ahead Market} 
The  DAM market-clearing problem minimizes the day-ahead system cost, which takes the following form:
\begin{subequations}\label{prob:DA}
\allowdisplaybreaks
\begin{align}
   \underset{\Phi^{\text{DA}}}{\min} ~&f^{\text{DA}}(\Phi^{\text{DA}}):= 
    \sum_{t\in\mathcal{T}}\sum_{k \in \mathcal{K}} \sum_{S \in \mathcal{S}} C_{k,t,s}^{\text{W}}  \cdot p_{k,t,s}^{\text{W}} \notag\\&\hspace{8ex}+
        \sum_{t\in\mathcal{T}}\sum_{i \in \mathcal{I}} (C_{i}  \cdot p_{i,t}^{\text{C}} +u_{i,t}^\text{DA}\cdot C_i^0+c^{\text{DA}}_{i,t} )\\
         \text{s.t.} 
	~&\sum_{i \in \mathcal{I}_{n}}p_{i,t}^{\text{C}} + \sum_{k \in \mathcal{K}_{n}}\sum_{s\in \mathcal{S}} p_{k,t,s}^{\text{W}} - L_{n,t} \notag\\ 
        &\hspace{1ex} -\sum_{m:(n,m)\in\Lambda}\frac{\delta_{n,t}^{\text{DA}}-\delta_{m,t}^{\text{DA}}}{x_{n,m}} = 0 
   :\lambda_{n,t}^b,\label{eq:dabalance}\\
	&-\overline{F}_{n,m}\leqslant \frac{\delta_{n,t}^{\text{DA}}-\delta_{m,t}^{\text{DA}}}{x_{n,m}} \leqslant \overline{F}_{n,m}:\underline{\lambda}_{n,m,t},\overline{\lambda}_{n,m,t},\label{eq:daline}\\
    &\pushright{\forall n\in\mathcal{N}, \forall (n,m) \in \Lambda,\forall t \in \mathcal{T},}\nonumber\\
    &0 \leqslant p_{k,t,s}^{\text{W}} \leqslant {{W}_{k,t,s}}:\underline{\lambda}_{k,t,s}^W, \overline{\lambda}_{k,t,s}^W, \label{eq:daw}\\
    &\pushright{\forall k\in \mathcal{K},\forall s\in \mathcal{S},\forall t \in \mathcal{T}},\nonumber\\
    &u_{i,t}^{\text{DA}}\cdot \underline{P}_{i}^{\text{C}} \leqslant p_{i,t}^{\text{C}} \leqslant u_{i,t}^{\text{DA}}\cdot \overline{P}_{i}^{\text{C}}:\underline{\lambda}_{i,t}^C, \overline{\lambda}_{i,t}^C,\label{eq:dac}\\
    & 0\leqslant u_{i,t}^{\text{DA}} \leqslant 1:\underline{\lambda}_{i,t}^U, \overline{\lambda}_{i,t}^U,\label{eq:dauc}\\
    & C_i^{\text{SU}}\cdot (u_{i,t}^{\text{DA}}-u_{i,t-1}^{\text{DA}})\leqslant c^{\text{DA}}_{i,t}:{\lambda}_{i,t}^{\text{S}_1},\label{eq:dasuc1}\\
    & 0\leqslant c^{\text{DA}}_{i,t}:{\lambda}_{i,t}^{\text{S}_2},\label{eq:dasuc2}\\
    & p_{i,t}^C-p_{i,t-1}^C \geqslant - u_{i,t-1}^{\text{DA}}\cdot R_i^\text{D} : \underline{\lambda}_{i,t}^R,   \label{eq:darp_a}\\
    &p_{i,t}^C-p_{i,t-1}^C \leqslant u_{i,t}^{\text{DA}}\cdot R_i^\text{U}: \overline{\lambda}_{i,t}^R,\label{eq:darp_b}\\
     &\pushright{\forall i \in  \mathcal{I}, \forall t \in \mathcal{T}}.\nonumber
	\end{align}
\end{subequations}
Here, the total dispatch cost $f^{\text{DA}}$ accounts for the variable, no-load, and startup costs of conventional generation units, along with the bidding costs of VRE producers. The decision variable set $\Phi^{\text{DA}}$ includes the day-ahead decisions, i.e., $p_{i,t}^{\text{C}}$ (generation output), $u_{i,t}^{\text{DA}}$ (commitment schedule), $c_{i,t}^\text{DA}$ (start-up cost), and  VRE generation at each bidding segment, i.e., $p_{k,t,s}^{\text{W}}$. Additionally, the set  includes voltage angles $\delta_{n,t}^{\text{DA}}$ for each bus. Constraint \eqref{eq:dabalance} ensures the power balance in the day-ahead market, while constraints \eqref{eq:daline} enforce power flow limits. Constraints \eqref{eq:daw} and \eqref{eq:dac} impose the generation limits of VRE and conventional generation units, respectively, with the latter accounting for the UC decisions. Lastly, constraints \eqref{eq:dauc}--\eqref{eq:darp_b} model the relaxed UC decisions, UC costs, and ramping limits. Here, the subscript $t=0$ denotes the initial state. 

Since we later focus on the bidding strategies of VRE, We denote by $\mathcal{X}^{\text{DA}}(\bm{W})$ the constraint set constructed by  \eqref{eq:dabalance}-\eqref{eq:darp_b} and parameterized by VRE bidding quantity $\bm{W}$ in the bidding curves. We assume that the true marginal cost of VRE is zero. Hence, we denote the actual DAM cost function  as $f_0^{\text{DA}}(\Phi^{\text{DA}})$, i.e., 
\begin{align}
 f_0^{\text{DA}}(\Phi^{\text{DA}}):= 
  \sum_{t\in\mathcal{T}}\sum_{i \in \mathcal{I}} C_{i}  \cdot p_{i,t}^{\text{C}} +u_{i,t}^\text{DA}\cdot C_i^0+c^{\text{DA}}_{i,t}.
\end{align}

\subsection{Real-Time Market}

Closer to real-time operations, any deviation from the day-ahead schedule $\Phi^{\text{DA}\star}$ is managed by balancing actions. For a specific realization $\omega\in \Omega$ of VRE generation {$\widetilde{W}_{k,\omega,t}$} and demand {${L}_{k,\omega,t}$}, the system operator determines the optimal re-dispatch by minimizing the re-dispatch cost $f_\omega^{\text{RT}}$. The cost includes the upward/downward adjustment  cost of conventional generators, startup cost and fixed operational cost  of new online generators, and the cost of lost load: 
\begin{subequations}\label{prob:RT}
\allowdisplaybreaks
	\begin{align}
&\underset{\Phi_\omega^{\text{RT}}}{\min}~  f_\omega^{\text{RT}}(\Phi_\omega^{\text{RT}},\Phi^{\text{DA}\star}):=\sum_{i \in \mathcal{I}}\sum_{t\in \mathcal{T}} \Big( C_{i}^U r_{i,t,\omega}^{\text{U}}- C_{i}^D r_{i,t,\omega}^{\text{D}} \notag \\
&+C_i^0\cdot (u_{i,t,\omega}^{\text{RT}}\!-u_{i,t}^{\text{DA}\star})\!+\!c_{i,t,\omega}^{\text{RT}}\Big) +\sum_{n \in \mathcal{N}}\sum_{t\in \mathcal{T}} C^{\text{sh}} l_{n,t,\omega}^{\text{sh}} \\
&	\text{s.t.}~  \sum_{i \in \mathcal{I}_{n}} \left( p_{i,t}^{\text{C}\star}+r_{i,t,\omega}^{\text{U}}-r_{i,t,\omega}^{\text{D}} \right)\!+\! \sum_{k \in \mathcal{K}_{n}}\!\left(\widetilde{W}_{k,t,\omega} -  p_{k,t,\omega}^{\text{W,cr}} \right) \notag  \\
&\hspace{6ex}- \hspace{-1ex}
\sum_{m:(n,m)\in\Lambda} \hspace{-2ex}\frac{\delta_{n,t,\omega}^{\text{RT}}-\delta_{m,t,\omega}^{\text{RT}}}{x_{n,m}} = L_{n,t,\omega}-l_{n,t,\omega}^{\text{sh}}, \label{eq:rtbalance}\\
& -\overline{F}_{n,m}\leqslant \frac{\delta_{n,t,\omega}^{\text{RT}}-\delta_{m,t,\omega}^{\text{RT}}}{x_{n,m}} \leqslant \overline{F}_{n,m},\label{eq:rtline} \\
&\pushright{\forall n \in \mathcal{N}, \forall (n,m) \in \Lambda, \forall t \in \mathcal{T}}\nonumber\\
& u_{i,t}^{\text{DA}\star} \leqslant u_{i,t,\omega}^{\text{RT}} \leqslant 1,\quad u_{j,t}^{\text{DA}\star} = u_{j,t,\omega}^{\text{RT}}, \label{eq:rtfs}\\
&\pushright{\forall i \in  \mathcal{I}^{\text{FS}}, \forall j \in  \mathcal{I}^{\text{SL}}, \forall t \in \mathcal{T},}\nonumber\\
& - u_{i,t,\omega}^{\text{RT}}\cdot \underline{P}_{i}\leqslant p_{i,t}^{\text{C}\star}+ r_{i,t,\omega}^{\text{U}}-r_{i,t,\omega}^{\text{D}} \leqslant  u_{i,t,\omega}^{\text{RT}}\cdot \overline{P}_{i}, \label{eq:rtup}\\
&c_{i,t,\omega}^{\text{RT}}+c_{i,t}^{\text{DA}\star}\geqslant C_i^{\text{SU}}\cdot (u_{i,t,\omega}^{\text{RT}}-u_{i,t-1,\omega}^{\text{RT}}),\label{eq:rtsuc1}\\
&-u_{i,t-1,\omega}^{\text{RT}}\cdot R_i^\text{D} \leqslant p_{i,t}^{\text{C}\star}+r_{i,t,\omega}^{\text{U}}-r_{i,t,\omega}^{\text{D}}\notag \\
&\hspace{6ex}-(p_{i,t-1}^{\text{C}\star}+r_{i,t-1,\omega}^{\text{U}}\!-\!r_{i,t-1,\omega}^{\text{D}}) \leqslant u_{i,t,\omega}^{\text{RT}}\!\cdot \!R_i^\text{U},\label{eq:rtrp}\\
&c_{i,t,\omega}^{\text{RT}}\geqslant 0,~r_{i,t,\omega}^{\text{U}}\geqslant0, ~r_{i,t,\omega}^{\text{D}}\geqslant0,\label{eq:rtsuc2}\\
&\pushright{\forall i \in \mathcal{I},\forall t \in \mathcal{T}}\\
& 0 \leqslant p_{k,t,\omega}^{\text{W,cr}} \leqslant \widetilde{W}_{k,t,\omega},\quad\quad\;\!\forall k \in \mathcal{K},\forall t \in \mathcal{T},\label{eq:rtw}\\
&0 \leqslant l_{n,t,\omega}^{\text{sh}} \leqslant L_{n,t,\omega}, \quad\quad\quad\!\!\forall n \in \mathcal{N},\forall t \in \mathcal{T}.\label{eq:rtlost}
\end{align}
\end{subequations}
The decision variable set $\Phi_\omega ^{\text{RT}}$ include the real-time re-dispatch variables of each conventional unit, i.e., $r_{i,t,\omega}^{\text{U}},r_{i,t,\omega}^{\text{D}}, u_{i,t,\omega}^{\text{RT}},c_{i,t,\omega}^{\text{RT}}$, curtailment of VRE generations, i.e., $p_{k,t,\omega}^{\text{W,cr}}$, and shed load $l_{n,t,\omega}^{\text{sh}}$. Additionally, it includes the real-time voltage angles $\delta_{n,t,\omega}^{\text{RT}}$. Constraint \eqref{eq:rtbalance} ensures the power balance in real time while constraint \eqref{eq:rtline} enforces transmission limits. 
The first entry in constraint \eqref{eq:rtfs} permits fast-start generators that were not committed in the DAM to be dispatched in the RTM. In contrast, the second entry in \eqref{eq:rtfs} ensures that the commitment status of slow-start generators in the RTM remains unchanged from their DAM commitment.  Constraint \eqref{eq:rtup} restricts the real-time generation of conventional generators, while \eqref{eq:rtsuc1} yields the startup costs in real time. Constraint \eqref{eq:rtrp} enforces ramping limits. Additionally, constraints \eqref{eq:rtw}, \eqref{eq:rtsuc2}, and \eqref{eq:rtlost} guarantee the feasibility of VRE curtailment, the upward/downward adjustment of conventional generators, and load shedding, respectively.

We denote by $\mathcal{X}_{\omega}^{\text{RT}}(\Phi ^{\text{DA}})$ the constraint set constructed by  \eqref{eq:rtbalance}-\eqref{eq:rtw}, which is parameterized by day-ahead schedule $\Phi ^{\text{DA}}$. 

			\tikzstyle{Clearing} = [rectangle, rounded corners = 5, minimum width=10, minimum height=10,text centered, draw=black, fill=white!30,line width=0.3mm]
   		\tikzstyle{Clearing1} = [rectangle, rounded corners = 5, minimum width=10, minimum height=10,text centered, draw=red, fill=white!30,line width=0.3mm]
			\tikzstyle{Reserve} = [rectangle, rounded corners = 5,  minimum width=10, minimum height=20,text centered, draw=black, fill=white!30,line width=0.3mm]

   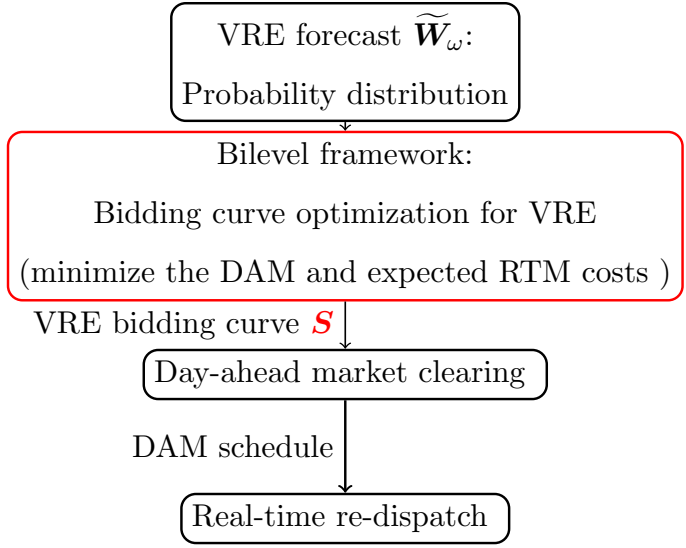
\begin{figure}[t]
   \centering
	        \begin{adjustbox}{width=90mm}
				\begin{tikzpicture}[node distance=50]
				
						\node [align=center] (Requirements_box) [Reserve] {VRE forecast $\widetilde{\bm{W}}_\omega$: \\Probability distribution};
						\node [align=center] (OR_box) [Clearing1, below of = Requirements_box, yshift = -3.5] {
		
						{Bilevel framework:}	\\	{Bidding curve optimization for VRE}  \\ {(minimize the DAM and expected RTM costs )}
						};
						
						\node [align=center] (DA_box) [Clearing, below of = OR_box, yshift = -5] {
							Day-ahead market clearing
						};

						\draw[transform canvas,->,line width=0.2mm] (OR_box) -- node[midway, left] {VRE bidding curve $\textcolor{red}{\bm{S}}$}  (DA_box);
						
						\draw[transform canvas={xshift=0cm},->,line width=0.2mm] (Requirements_box) -- 
						node[midway, left, align=center] {}
						(OR_box) ;
						\node [align=center] (RT_box) [Clearing, below of=DA_box, xshift=0] {
							Real-time re-dispatch  
						};
						\draw[transform canvas={xshift=0cm},->,line width=0.3mm] (DA_box) -- node[midway, left] {DAM
      schedule}  (RT_box) ;

				\end{tikzpicture}

\end{adjustbox}
\caption{Bid optimization and market-clearing timeline}
\label{fig:market}
\vspace{-2ex}
\end{figure}

\section{A General Bilevel Framework}\label{section:bilevel}

Conventionally, VRE producers bid zero cost with the day-ahead forecast values in the DAM. However, this strategy cannot account for potential re-dispatch costs caused by VRE uncertainty, which cannot guarantee market efficiency.  Building on the sequential DAM \eqref{prob:DA}  and RTM \eqref{prob:RT} problems,  this work develops a bilevel stochastic framework (\textit{BiD}) to reveal the optimal bidding curves of VRE producers to improve the economic coordination between the two markets.

As illustrated in Fig. \ref{fig:market}, we introduce this framework ahead of the DAM  clearing. The obtained optimal bidding curves of VRE in the bilevel framework will be incorporated into the DAM scheduling. The RTM is then cleared accordingly. Notably, this bilevel model will not change the sequential DAM and RTM clearing structure.

Now, we go into the detail of the bilevel framework \textit{BiD}, which is based on a bilevel optimization problem. In the upper level, the system operator decides the day-ahead VRE bidding curves $\mathcal{S}$, i.e., the price and quantity pair $(\bm{C}^{\text{W}},\bm{W})$. In the lower level, the DAM schedule $\Phi^{\text{DA}}$ is optimized under the VRE bidding curves $\mathcal{S}$ provided from the upper  level. Sourcing the optimal DAM dispatch $\Phi^{\text{DA}}$ from the lower level, the system operator in the upper level also optimizes the RTM re-dispatch $\Phi_{\omega}^{\text{RT}}$ for different uncertainty realizations, so as to minimize the expected cost. The overall goal is to minimize the total system cost, i.e. the sum of DAM and expected RTM costs. The bilevel problem takes the following form:

\begin{tcolorbox}[standard jigsaw,opacityback=0]
\vspace{-1ex}
\textbf{Problem \textit{BiD}: Bilevel optimization problem for day-ahead VRE bidding curves}
\begin{subequations} \label{prob:bilevel_clearing}	
\allowdisplaybreaks
	\begin{align*}
\hspace{-2.5ex}S^{\text{BiD}}:=\underset{\Phi^{\text{RT}}\bigcup (\bm{C}^{\text{W}},\bm{W})}{\text{min}}& ~ \!
		f_0^{\text{DA}}(\Phi^{\text{DA}\star}) + \mathbb{E}_{\omega\in \Omega}\left[f_\omega^{\text{RT}}(\Phi_\omega^{\text{RT}},\Phi_\omega^{\text{DA}\star})\right] \\	
& \hspace{-7ex}\text{s.t.}~  \Phi_\omega^{\text{RT}}\in \mathcal{X}_\omega^{\text{RT}}(\Phi^{\text{DA}}),~\forall \omega \in \Omega,\\
		&\hspace{-3ex} \Phi^{\text{DA}\star}  \in \text{arg}
		\left\{\!\begin{aligned}
			\underset{\Phi^{\text{DA}}}{\text{min}} ~&f^{\text{DA}}(\Phi^{\text{DA}})\\
			\text{s.t.} ~& \Phi^{\text{DA}}\in \mathcal{X}^{\text{DA}}(\bm{W})\\
		\end{aligned}\right\}. 
	\end{align*}
\end{subequations}
\end{tcolorbox}

Once the optimal bid curve $(\bm{C}^{\text{W*}},\bm{W}^\star)$ is obtained from Problem \textit{BiD}, the system operator first clears the day-ahead market \eqref{prob:DA} using $(\bm{C}^{\text{W*}},\bm{W}^\star)$, and then, closer to real-time, clears the RTM \eqref{prob:RT} for a particular realization of renewable generation. The expected system cost is denoted by $S^{\text{BiD}}$.

Note that this bilevel framework focuses on the system perspective to establish benchmark bidding curves for VRE in a centralized manner. We expect that this bilevel framework can serve as a tool to guide, monitor, or regulate the bidding strategies of private suppliers. For example, by comparing the benchmark results with
the forecast values, a risk score can be developed as a coordination mechanism to adjust and guide how much suppliers should
bid under the bidding prices \citep{sur2024application}. Another interesting future direction is to  develop incentive mechanisms to guide VRE producers' bidding strategies, ensuring alignment with the optimal outcomes of the centralized framework.


\section{A Single-Bidding-Segment Model Achieving Optimality} \label{section:equiv}

Problem \textit{BiD} is challenging to solve when jointly optimizing the bidding price and bidding quantity for each segment, as the number of corresponding bilinear terms in the objective function increases in the number of segments. However, we find that a simplified  case can guarantee optimality.


We consider a simplified case of bidding curves as follows: \textit{There is but one segment of the bidding curve and it has a zero price, i.e.,  $C_1=0$ \$/MWh for all the hours and VRE producers.}  Note that zero is the true marginal cost for VRE. The bilevel problem then only optimizes the quantity of that single segment. We denote this bilevel optimization as Problem  \textit{BiD-q}, which is presented in detail as follows. 

\begin{tcolorbox}[standard jigsaw,opacityback=0]
\textbf{Problem \textit{BiD-q}: Bilevel optimization problem for quantity-only adjustment}
\begin{subequations} \label{prob:bilevel_clearingx}	
\allowdisplaybreaks
	\begin{align*}
\underset{\Phi^{\text{RT}}\bigcup \bm{W}}{\text{min}}& ~ \!
		f_0^{\text{DA}}(\Phi^{\text{DA}\star}) + \mathbb{E}_{\omega\in \Omega}\left[f_\omega^{\text{RT}}(\Phi_\omega^{\text{RT}},\Phi_\omega^{\text{DA}\star})\right] \\	
& \hspace{-7ex}\text{s.t.}~  \Phi_\omega^{\text{RT}}\in \mathcal{X}_\omega^{\text{RT}}(\Phi^{\text{DA}}),~\forall \omega \in \Omega,\\
		&\hspace{-3ex} \Phi^{\text{DA}\star}  \in \text{arg}
		\left\{\!\begin{aligned}
			\underset{\Phi^{\text{DA}}}{\text{min}} ~&f_0^{\text{DA}}(\Phi^{\text{DA}})\\
			\text{s.t.} ~& \Phi^{\text{DA}}\in \mathcal{X}^{\text{DA}}(\bm{W})\\
		\end{aligned}\right\}. 
	\end{align*}
\end{subequations}
\end{tcolorbox}

 The next result shows that from the system perspective, the quantity-only adjustment under zero bidding prices achieves the same expected system cost as the original Problem \textit{BiD}.
\begin{thm}\label{prop:same}
  Problem \textit{BiD} and Problem  \text{BiD-q} achieve the same expected system cost.
\end{thm}

We present the proof in the appendix. Next, we posit the following corollary.
\begin{coro}\label{rem}
Considering multi-segment bidding curves, i.e., $S\geqslant 2$, as long as one segment is set by a zero price, the bilevel framework will achieve the same expected system cost as the original Problem \textit{BiD}.
\end{coro}

However, if the bidding prices are not zeros, the expected system cost under the bilevel framework may be higher. For example, in the extreme case of the very expensive single-segment bidding curve, no VRE will be dispatched in the DAM, which is typically not cost-optimal for the system. 

In the next section, we leverage these theoretical results in the interest of numerical experiments. Since only zero price is needed to ensure optimality, we will simulate the results of the bilevel framework when the bidding prices are given in the bidding curves.

\section{Case study } \label{section:givenprice}

VRE producers in practice are allowed to bid multi-segment curves with non-zero prices. Here, we test the bilevel framework for single- and multi-segment bidding curves, respectively, \textit{for the case where bidding prices are fixed inputs.} For simplicity, we still call those bilevel optimization models as Problem \textit{BiD}. Before introducing simulation results, we present our solution method of solving the bilevel model for large-scale systems.

\subsection{Solution Method}
When the bidding price is given, the upper- and lower-level formulations in Problem \textit{BiD} are LPs. The conventional method is to substitute the lower-level problem with its KKT conditions and leverage the Big-M reformulation of complementarity slackness\citep{morales2014electricity}. While this method works well on a small-scale system (e.g., IEEE 118-bus system), it does not scale well for real-life applications (e.g., NYISO) \citep{zhao2022uncertainty}. 

As a solution,  our prior work developed a method based on the strong duality and McCormick-envelope relaxation \citep{zhao2022uncertainty}. With the KKT conditions of the lower-level problem, the idea is to first substitute the complementarity slackness constraints with the strong duality condition, and then address the resulting bilinear terms, i.e., $\overline{\lambda}_{k,t,s}^W \cdot W_{k,t,s}$, with McCormick envelopes. In this way, we relax the  KKT conditions of the lower-level problem into linear constraints, and thus relax the whole bilevel optimization problem into an LP. The new LP can be efficiently solved at the scale of the 1576-bus NYISO system. The details can be found in the paper \citep{zhao2022uncertainty}. 

\subsection{NYISO System} The tested  NYISO system we use has 1576 buses, 2359 transmission lines, 1564 loads, 345 conventional generation units,  and 27 wind farms \citep{greene2022}. We use the wind and load data of August 2, 2019. We generate 20 joint scenarios for the probability distribution of wind and load forecasts using PGscen \citep{carmona2022joint}. We increase the wind-energy capacity so that its average generation amounts to 40\% of the total demand.

 For illustration, we adopt a short operation window, e.g.,  from 7:00 am to 10:00 am, as the simulations on 24-hour operations and a large number of scenarios are computationally challenging.  The 24-hour schedule results can be approximated through the computation over 4-hour rolling horizons.

\subsection{Benchmarks}
 We consider two benchmarks for the Problem \textit{BiD}: [1] \textit{Myopic dispatch (MyD):} Each VRE producer $k$ offers the expected value of the forecast for a one-segment bidding curve, i.e.,   ${W}_{k,t}=\mathbb{E}_{\omega \in \Omega}[\widetilde{W}_{k,t,\omega}]$ in \eqref{eq:daw}.   We denote the expected system cost as $S^{\text{MyD}}$ including the costs of DAM and RTM. [2] \textit{Stochastic dispatch (StD)}: Stochastic co-optimization of the DAM and RTM schedules by minimizing the total expected costs across. Although least-cost, this solution is not compatible with existing market designs.
In terms of system costs under  \textit{StD}, \textit{BiD} and \textit{MyD}, the system costs  always satisfy  $S^{\text{MyD}}\geqslant S^{\text{BiD}}\geqslant S^{\text{StD}}$ \citep{zhao2022uncertainty}.

\subsection{Single-Segment Biding Curves} 

We now consider the single-segment bidding curves with non-negative bidding prices. Since zero bidding prices can already achieve the system optimum (see Corollary \ref{rem}), we will vary the bidding price to demonstrate its impact.

\begin{figure}[t]
	\centering
	\hspace{-1ex}
	\subfigure[]{
		\raisebox{-1mm}{\includegraphics[width=2.3in]{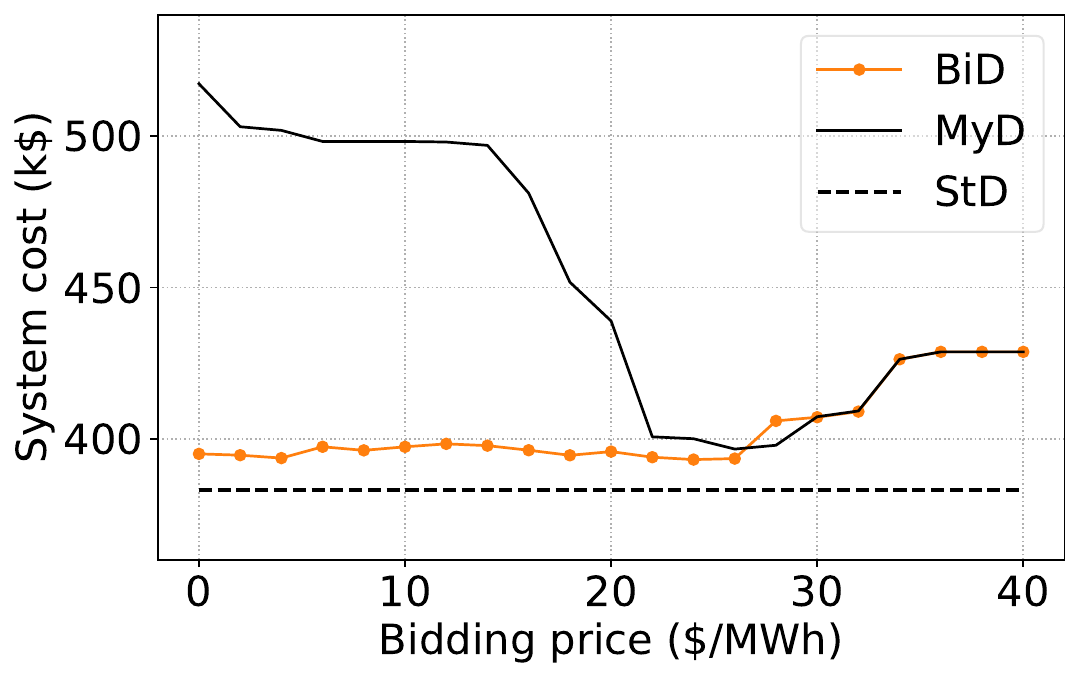}}}
	\hspace{-1ex}
	\subfigure[]{
		\raisebox{-1mm}{\includegraphics[width=2.3in]{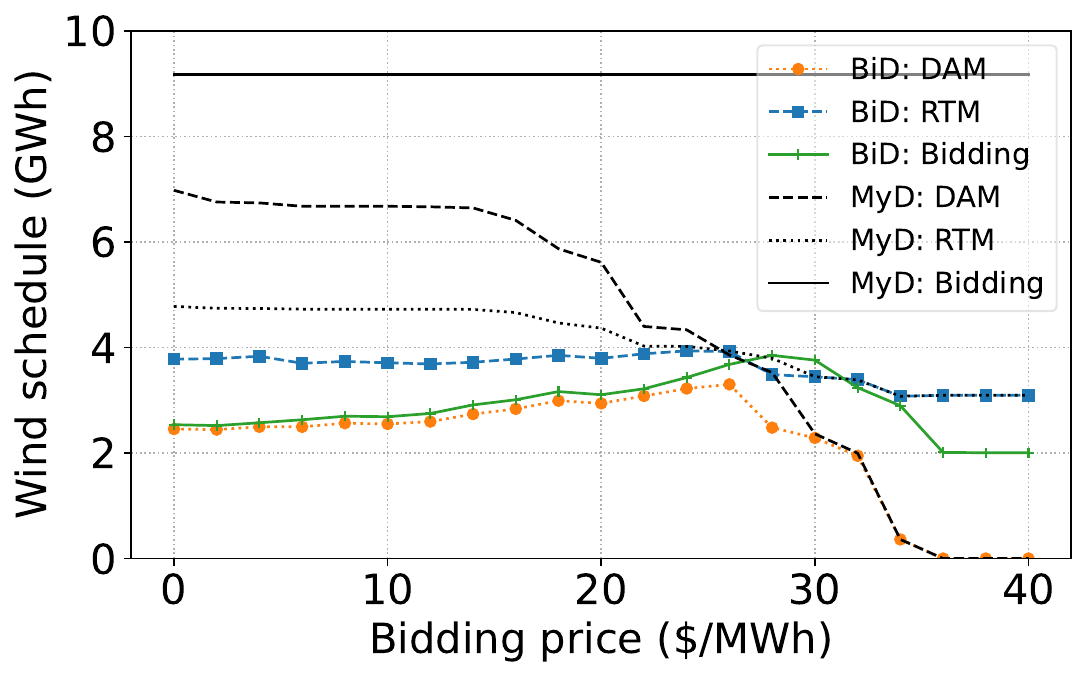}}}
  \hspace{-1ex}
  	\subfigure[]{
		\raisebox{-1mm}{\includegraphics[width=2.3in]{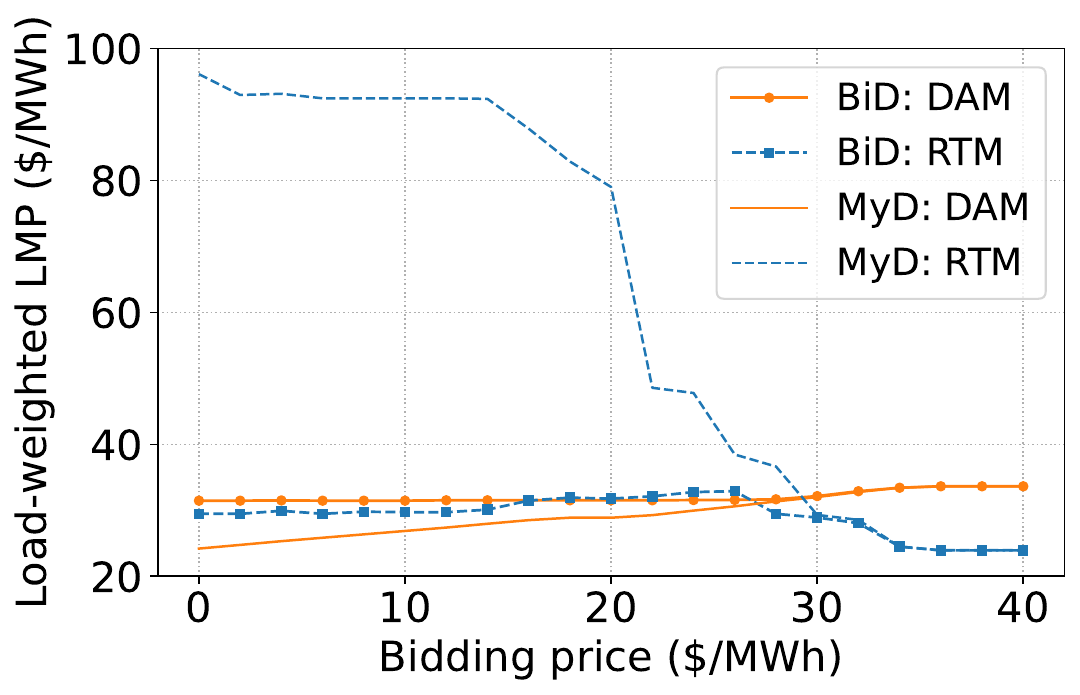}}}
  	\hspace{-1ex}
  	\subfigure[]{
		\raisebox{-1mm}{\includegraphics[width=2.3in]{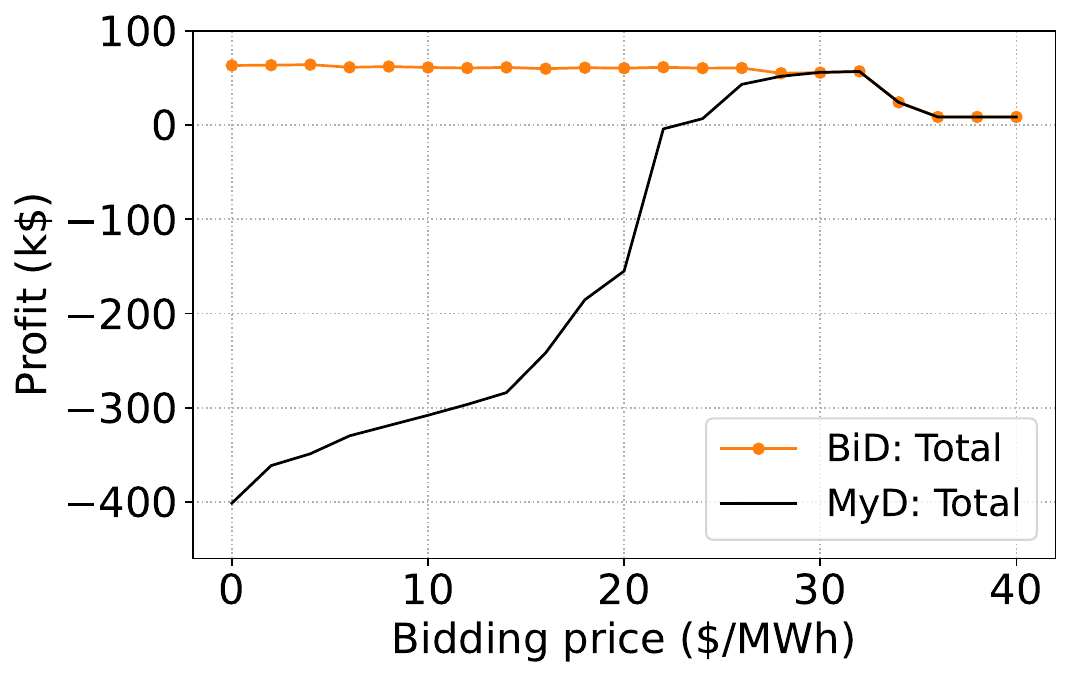}}}
  \vspace{-1ex}
	\caption{ Results (per hour) of \textit{BiD} vs \textit{MyD}: (a)  System cost; (b) Wind schedule amount; (c) Load-weighted LMP; (d) Aggregate profits of all wind farms.}
	\label{fig:single}
 \vspace{-2ex}
\end{figure}

Fig. \ref{fig:single} shows that under the bilevel framework, varying bidding prices within a certain range will not significantly change the system cost and wind producers' aggregate profits due to the dynamic bidding quantity adjustment and conservative bidding in the DAM. However, under \textit{MyD}, the bidding price can have significant cost and VRE profit impacts. 
Specifically,  Fig. \ref{fig:single}(a) shows that when the bidding price is lower than 26\$/MWh, the system cost under \textit{BiD} does not change much, and  is close to the least-cost benchmark \textit{StD}. However, the system cost under \textit{MyD} can be much higher than \textit{BiD}. Note that when the bidding price is low,  \textit{MyD} tends to bid much more wind energy than \textit{BiD} as shown in Fig. \ref{fig:single}(b). This poses a higher shortage risk in the RTM and leads to a high system cost. Thus, the increasing bidding price will reduce the DAM wind schedule amount under \textit{MyD} and reduce the system cost. However, when the bidding price is higher than 26\$/MWh, the system cost will increase for both \textit{BiD} and \textit{MyD}. After the 36\$/MWh price threshold, no wind will be dispatched in the DAM for both \textit{BiD} and \textit{MyD} as shown in Fig. \ref{fig:single}(b), which leads to the same high system cost for both \textit{BiD} and \textit{MyD}.

Fig. \ref{fig:single}(c) further reveals that under \textit{BiD}, the DAM and RTM prices do not change much when the bidding price increases. When bidding higher than 26 \$/MWh, the DAM prices will increase as less wind energy is dispatched in the DAM. The RTM prices decrease since less wind energy in the DAM yields lower shortage risks in the RTM. In contrast, under \textit{MyD}, the DAM prices and RTM prices vary significantly with an increasing bidding price. Accordingly, in Fig. \ref{fig:single}(d), under \textit{BiD}, the expected aggregate profit of wind producers remains almost the same when the bidding price is lower than 26 \$/MWh. The profit sees a drop when the bidding price approaches 36 \$/MWh because fewer revenues are collected in the DAM. However, under \textit{MyD}, the profit can be very low at negative values when the bidding price is low, which is due to the high shortage payment in the RTM.

\subsection{Multi-Segment Bidding Curves}

Although the single-segment zero-price bid achieves the system optimum, VRE producers can bid multi-segment curves in practice. We demonstrate that the proposed bilevel model can be used to set up the benchmark bidding quantity for each segment so as to approximate the system optimum.

For the bidding curve settings, we consider the case when the zero price is included in the bidding curve.
 We empirically set prices for 6-segment bidding curves, which are denoted by $0=C_1^{\text{W}}<C_2^{\text{W}},\ldots,<C_6^{\text{W}}$. We examine the supply curve for all conventional generations based on the operational cost and accumulative capacity. We set $C_1^{\text{W}}=0 \$ $/MWh, $C_2^{\text{W}}=2 \$ $/MWh, $C_3^{\text{W}}=22 \$ $/MWh, $C_4^{\text{W}}=30 \$ $/MWh, $C_5^{\text{W}}=32 \$ $/MWh,  $C_6^{\text{W}}=350 \$ $/MWh.  $C_2^{\text{W}}$ to $C_5^{\text{W}}$ are set based on the costs of the first 20\%, 40\%, 60\%, and 80\% of the total conventional capacity. $C_6^{\text{W}}$ is set at a price higher than the maximum cost of the exiting units.  We apply the \textit{BiD} framework to compute the bidding quantity  $W_{k,s,t} $ for each wind producer $k$, segment $s$, and time $t$. 

We will demonstrate optimal bidding curves from our bilevel model and evaluate the system cost.

 \begin{figure}[t]
	\centering
	\hspace{-2ex}
	\subfigure[]{
		\raisebox{-1mm}{\includegraphics[width=2.3in]{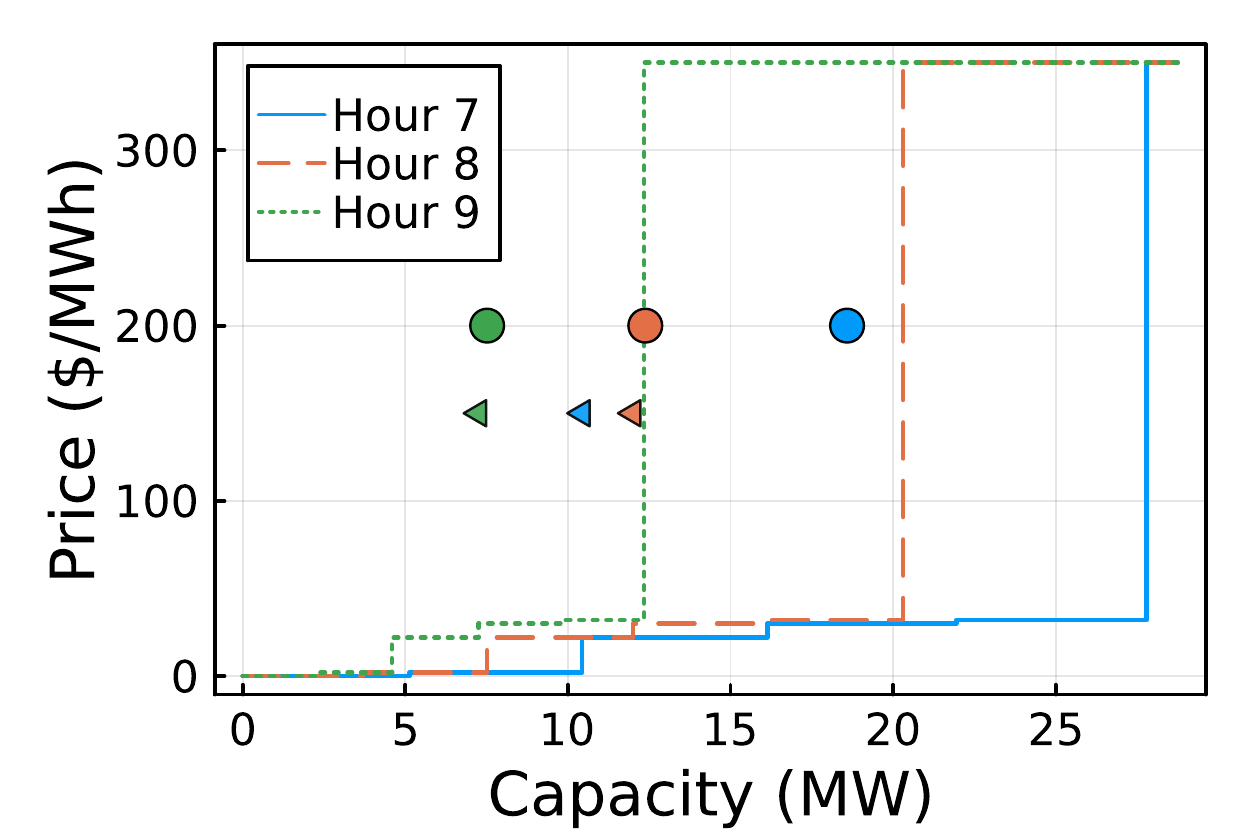}}}
	\hspace{-1ex}
	\subfigure[]{
		\raisebox{-1mm}{\includegraphics[width=2.3in]{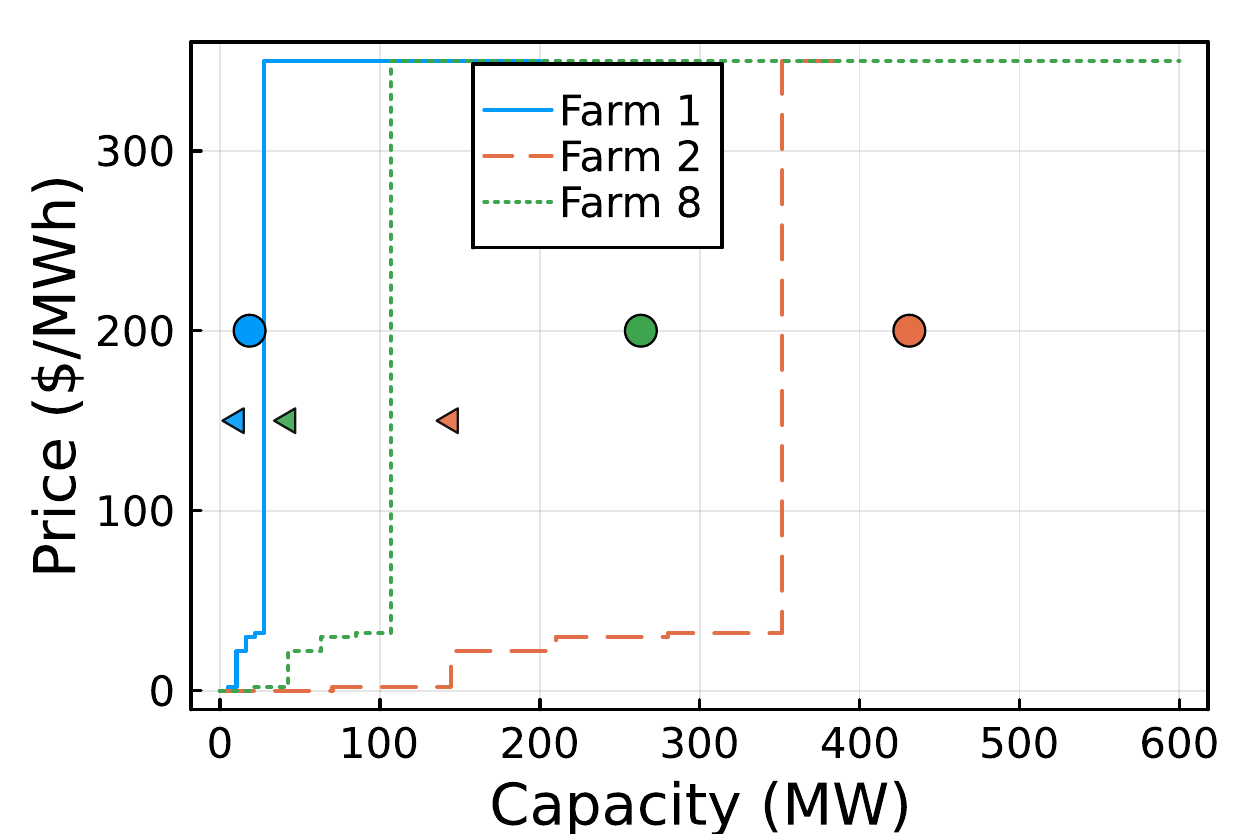}}}
	\caption{ (a) Wind farm 1 at hours 7 (blue), 8 (red), and 9 (green). (b)  Wind farms 1 (blue), 2 (red), and 8 (green) at hour 7.  The circle markers show the expected generation amount. The triangle markers show the DAM schedule. }
	\label{fig:biddingcurve}
 \vspace{-2ex}
\end{figure} 

 \begin{figure}[t]
	\centering
	\hspace{-2ex}
	\subfigure[]{
		\raisebox{-1mm}{\includegraphics[width=2.3in]{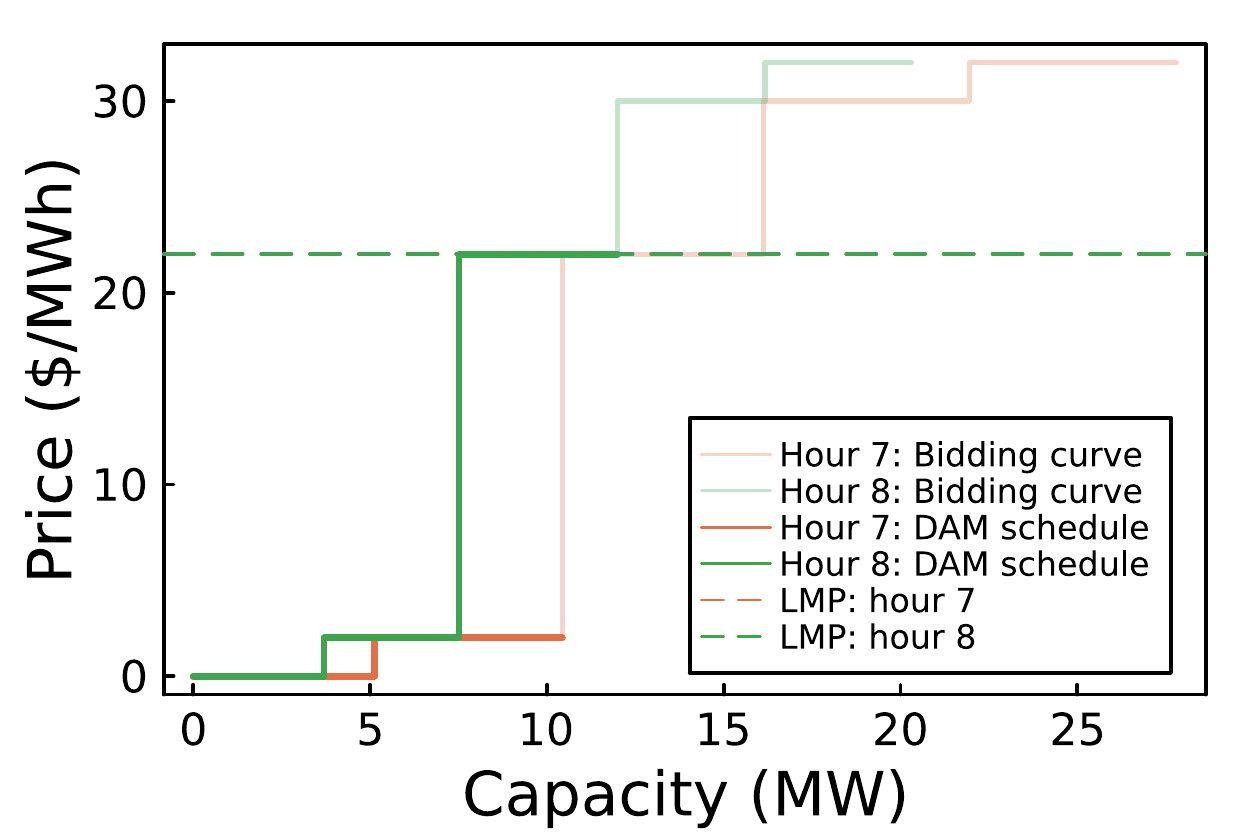}}}
	\hspace{-2ex}
	\subfigure[]{
		\raisebox{-1mm}{\includegraphics[width=2.3in]{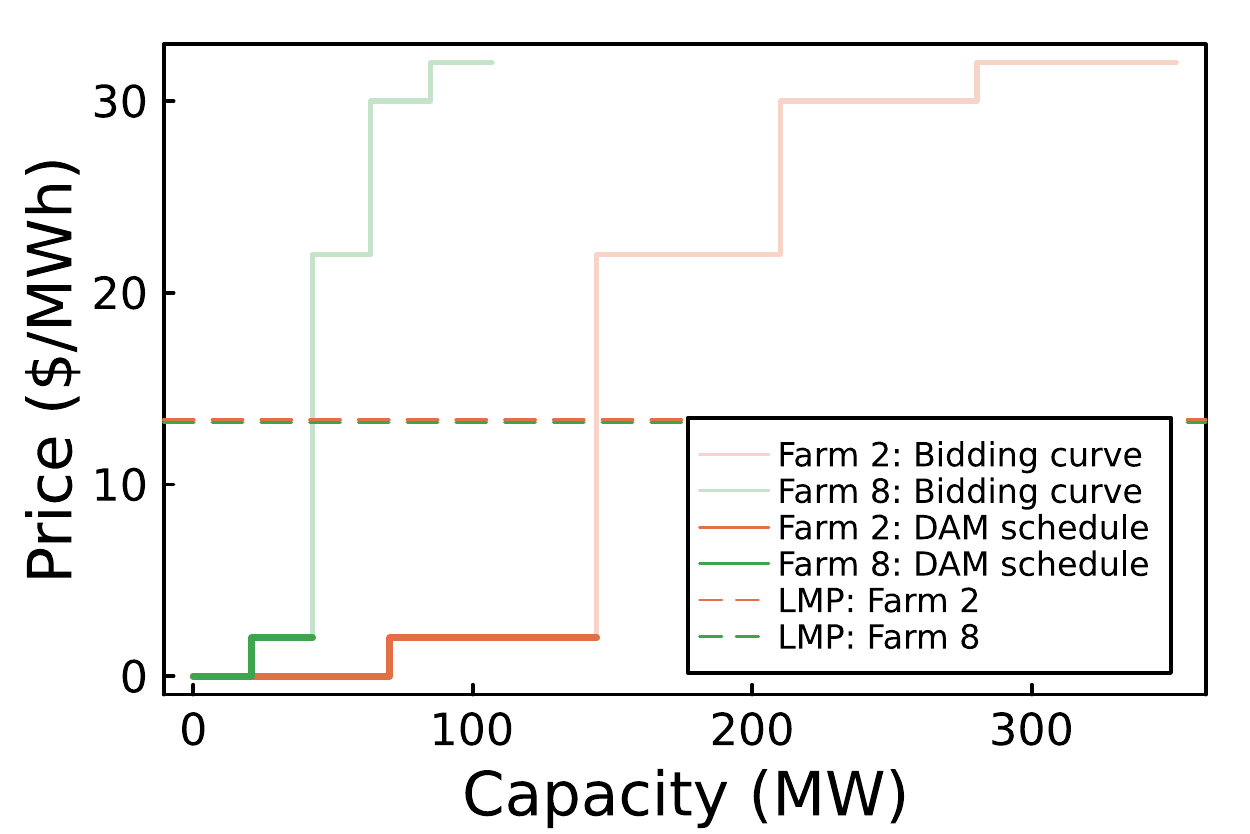}}}
	\caption{  DAM schedule (highlighted-color solid curves) vs. bidding curves (light-color solid curves).    (a) Wind farm 1 at hours 8 (red) and 9 (green);  (b) Wind farms 2 (red) and 8 (green) at hour 7. The horizontal curves show the LMPs of the buses where the wind farm is located.   }
	\label{fig:biddingandschdule}
 \vspace{-2ex}
\end{figure} 

\textit{Bidding curves: } 
Fig. \ref{fig:biddingcurve} displays the optimized bidding curves of selected wind farms at selected hours. 
The optimal curves, which vary for different hours and wind farms, are significantly affected by the expected wind power forecasts (in circles). As expected, a higher forecast amount can lead to more right-shifted bidding curves, and thus the DAM scheduled amount of wind energy (in triangles) may be higher.

\textit{DAM scheduled wind energy based on bidding curves:}  
Fig. \ref{fig:biddingandschdule} reveals how much wind energy gets dispatched under the optimal bidding curves. 
It is observed that, although we construct multiple segments, usually the first three segments will be dispatched in the DAM. 
Also, as shown by the horizontal curves in Fig. \ref{fig:biddingandschdule}(a) and (b), the bidding curves of wind energy may or may not set the LMPs at its located bus.

 \textit{System cost evaluations}: 
Finally, we report the following expected dispatch costs:
\begin{align*}
    S^{\text{MyD}}=\$430k,\quad S^{\text{BiD}}=\$275k,\quad S^{\text{StD}}=\$263k
\end{align*}
where the multi-segment optimized bid demonstrates a significant cost-saving potential with respect to the myopic strategy, and efficiently approximates the stochastic dispatch solution. The single-segment curves with zero price report the cost at $S^{\text{BiD}}=\$276k$, which is almost the same as the above case of multiple segments.\footnote{Although theoretically the multi- and single-segment curves achieve the same system cost, the simulation results can have slight differences.}


%

\section{Conclusion} \label{section:conclusion} 
This work formulates a bilevel optimization framework to optimize DAM bidding curves of VRE producers. The proposed framework internalizes the cost of real-time re-dispatch into day-ahead bidding curves, resulting in the minimum expected dispatch costs across the market timeline. Although the bilevel model is challenging to solve, we prove that from the system perspective, the single-segment bidding curve with zero bidding price is sufficient to achieve optimality if the true marginal cost of VRE is zero. 
We test the bilevel framework for single- and multi-segment bidding curves at scale on the 1576-bus NYISO system, revealing the impacts of DAM bidding curves on the overall market outcomes. 
Specifically, the optimized biding curves allow for a 36\% $(-\$155k)$ reduction in hourly system costs with respect to the baseline, where VRE producers bid their expected forecasts. Moreover, this cost-saving result is very close to that under the scenario-based stochastic market clearing, which provides the ideal coordination yet remains incompatible with existing market practices. Overall, our proposed framework can inform system-optimal VRE bidding strategies. Moreover, it can potentially serve as a tool to guide, monitor, or regulate VRE bids, e.g., through risk scores \citep{sur2024application}.

Our current model does not account for energy storage and demand response. These resources can be crucial for supporting deeply decarbonized electricity markets and we plan to incorporate these aspects in future work. Additionally, our existing bilevel optimization framework employs a centralized approach. In the future, we aim to develop incentive mechanisms to guide VRE producers' bidding strategies, ensuring alignment with the optimal outcomes of the centralized framework.

\section{Acknowledgment}
The information, data, or work presented herein was funded by the Advanced Research Projects Agency-Energy (ARPA-E), U.S. Department of Energy, under Award Number DE-AR0001277. The views and opinions of authors expressed herein do not necessarily state or reflect those of the United States Government or any agency thereof.


\appendix
\section{Proof of Theorem \ref{prop:same}}
Since Problem \textit{BiD-q} is a special case of Problem \textit{BiD}, we always have $S^{\text{BiD}}\leqslant S^{\text{BiD-q}}$. We thus only need to prove $S^{\text{BiD}}\geqslant S^{\text{BiD-q}}$ to arrive at equality.
The idea is to show that given the optimal solution to Problem \textit{BiD}, we can always find a feasible solution to Problem \textit{BiD-q}, which achieves the same system cost as  Problem \textit{BiD}.

First, we define some notations for the optimal solution to  Problem \textit{BiD}.  We denote the optimal bidding curve as $(\bm{C}^{\text{W}\star},\bm{W}^\star)$, the optimal DAM dispatch solution as $\Phi^{\text{DA}\star}$. The latter includes VRE schedule $p^{\text{W}\star}_{k,t,s}$ for any $k,t,s$. We denote the DAM dispatch solution, except VRE schedule $\bm{p}^{\text{W}\star}=(p^{\text{W}\star}_{k,t,s}, \forall k,t,s)$, as $\Phi_{-W}^{\text{DA}\star}$, i.e., $\Phi^{\text{DA}\star}=(\bm{p}^{\text{W}\star},\Phi_{-W}^{\text{DA}\star})$.

Second, we construct feasible solutions for  Problem \textit{BiD-q} based on the optimal solution to Problem \textit{BiD}.  Note that the bidding prices satisfy $C_{k,t,1}^\star\leqslant C_{k,t,2}^\star,...,\leqslant C_{k,t,S}^\star$. Thus,  in terms of optimal solutions to Problem \textit{BiD},  there exists an $s'$ such that for any $1\leqslant s \leqslant s'$, $p^{\text{W}*}_{k,t,s}\geqslant 0$, and for any $s'+1 \leqslant s \leqslant S$, $p^{\text{W}*}_{k,t,s}=0$. We now construct feasible solutions   $\bm{W}^\dagger$ and $\Phi^{\text{DA}\dagger}$ for Problem \textit{BiD-q}. We let 
\begin{align}
&W_{k,t}^\dagger=p_{k,t}^{\text{W}\dagger}=\sum_{1\leqslant s\leqslant s'}p^{\text{W}\star}_{k,t,s},\label{eq:proofint2}\\
&\Phi_{-W}^{\text{DA}\dagger}=\Phi_{-W}^{\text{DA}\star}.\label{eq:proofint3}
\end{align} The constructed solutions  $\bm{W}^\dagger$ and $\Phi^{\text{DA}\dagger} =\left(\bm{p}^{\text{W}\dagger},\Phi_{-W}^{\text{DA}\dagger}\right) $ satisfy the constraint $\Phi^{\text{DA}}\in \mathcal{X}^{\text{DA}}(\bm{W}^\dagger )$ in the lower-level problem of Problem \textit{BiD-q}.

Third, we prove by contradiction that $\Phi^{\text{DA}\dagger} =\left(\bm{p}^{\text{W}\dagger},\Phi_{-W}^{\text{DA}\dagger}\right) $ is an optimal solution to the following lower-level problem under the bidding quantity $\bm{W}^\dagger$ in Problem \textit{BiD-q}.
\begin{subequations}\label{eq:prooflowerlevel}
\begin{align}
\underset{\Phi^{\text{DA}}}{\text{min}} ~&f_0^{\text{DA}}(\Phi^{\text{DA}})\\
\text{s.t.} ~& \Phi^{\text{DA}}\in \mathcal{X}^{\text{DA}}(\bm{W}^\dagger ).
\end{align}
\end{subequations}

To begin with, we assume that the optimal DAM dispatch solution to the above problem \eqref{eq:prooflowerlevel} is $\Phi^{\text{DA}\ddagger}=\left(\bm{p}^{\text{W}\ddagger},\Phi_{-W}^{\text{DA}\ddagger}\right)$. Note that ${p}_{k,t}^{\text{W}\ddagger}\leqslant {W}_{k,t}^\dagger={p}_{k,t}^{\text{W}\dagger}$. If ${p}_{k,t}^{\text{W}\ddagger}= {p}_{k,t}^{\text{W}\dagger}$,  we easily have $\Phi_{-W}^{\text{DA}\ddagger}=\Phi_{-W}^{\text{DA}\dagger}$. 

Then, suppose $\left(\bm{p}^{\text{W}\dagger},\Phi_{-W}^{\text{DA}\dagger}\right) $  is not an optimal solution to the problem \eqref{eq:prooflowerlevel}, i.e.,  we can assume that there exist $k_\alpha$ and $t_\alpha$ such that $p_{k_\alpha,t_\alpha}^{\text{W}\ddagger}<p_{k_\alpha,t_\alpha}^{\text{W}\dagger}={W}_{k,t}^\dagger$. Thus, we have 
\begin{align}
f_0^{\text{DA}}(\Phi^{\text{DA}\ddagger})< f_0^{\text{DA}}(\Phi^{\text{DA}\dagger})=f_0^{\text{DA}}(\Phi^{\text{DA}\star}). \label{eq:proofint1}
\end{align}
The latter equality in \eqref{eq:proofint1}  is because  $f_0^{\text{DA}}$ is only related to $\Phi_{-W}^{\text{DA}}$ and  we have $\Phi_{-W}^{\text{DA}\dagger}=\Phi_{-W}^{\text{DA}\star}$.

Now, back to Problem \textit{BiD},  there will exist  $\mathcal{S}_\alpha$ and $\tilde{p}_{k,t,s}^{\text{W}\ddagger}$ such that (i) $\tilde{p}_{k_\alpha,t_\alpha,s_\alpha}^{\text{W}\ddagger}<p_{k_\alpha,t_\alpha,s_\alpha}^{\text{W}\star}$, $\forall s_\alpha \in \mathcal{S}_\alpha$; (ii) If $k\neq k_\alpha $  or $t\neq t_\alpha $ or $s \notin \mathcal{S}_\alpha $, $\tilde{p}_{k,t,s}^{\text{W}\ddagger}=p_{k,t,s}^{\text{W}\star}$; and (iii) $\sum_{1\leqslant s\leqslant s'}\tilde{p}^{\text{W}\ddagger}_{k,t,s}=p_{k,t}^{\text{W}\ddagger}$. Thus, we have a new feasible solution $\tilde{\Phi}^{\text{DA}\ddagger} =\left(\tilde{\bm{p}}^{\text{W}\ddagger},\Phi_{-W}^{\text{DA}\ddagger}\right) $ satisfying the lower problem constraint $\mathcal{X}^{\text{DA}}(\bm{W}^\star )$ under Problem \textit{BiD}. Based on (i), (ii), and \eqref{eq:proofint1}, we have 
\begin{align*}
&f_0^{\text{DA}}(\tilde{\Phi}^{\text{DA}\ddagger})+\sum_{t\in\mathcal{T}}\sum_{k \in \mathcal{K}} \sum_{S \in \mathcal{S}} C_{k,t,s}^{\text{W}\star}  \cdot p_{k,s,t}^{\text{W}\ddagger} \\
&\hspace{6ex}< f_0^{\text{DA}}(\Phi^{\text{DA}\star})+\sum_{t\in\mathcal{T}}\sum_{k \in \mathcal{K}} \sum_{S \in \mathcal{S}} C_{k,t,s}^{\text{W}\star}  \cdot p_{k,t,s}^{\text{W}\star}. \notag  
\end{align*}
This contradicts the fact that $\Phi^{\text{DA}\star}$ is the optimal solution to 
\begin{align*}
\underset{\Phi^{\text{DA}}}{\text{min}} ~&f^{\text{DA}}(\Phi^{\text{DA}})\\
\text{s.t.} ~& \Phi^{\text{DA}}\in \mathcal{X}^{\text{DA}}(\bm{W}^\star ).
\end{align*}

Therefore,   $\left(\bm{p}^{\text{W}\dagger},\Phi_{-W}^{\text{DA}\dagger}\right) $  is an optimal solution to the problem \eqref{eq:prooflowerlevel}, i.e., $\left(\bm{p}^{\text{W}\dagger},\Phi_{-W}^{\text{DA}\dagger}\right) $ is feasible to Problem \textit{BiD-q}. Note that we have $f_0^{\text{DA}}(\Phi^{\text{DA}\dagger})=f_0^{\text{DA}}(\Phi^{\text{DA}\star})$ shown in \eqref{eq:proofint1}.  Based on \eqref{eq:proofint2} and \eqref{eq:proofint3}, the constraint $\Phi_\omega^{\text{RT}}\in \mathcal{X}_\omega^{\text{RT}}(\Phi^{\text{DA}\dagger})$ of Problem  \textit{BiD-q} will be the same as $\Phi_\omega^{\text{RT}}\in \mathcal{X}_\omega^{\text{RT}}(\Phi^{\text{DA}\star})$ of Problem  \textit{BiD}. Therefore,  the objective, i.e., the expected system cost, of Problem  \textit{BiD-q} under  $\left(\bm{p}^{\text{W}\dagger},\Phi_{-W}^{\text{DA}\dagger}\right) $ is equal to that of Problem  \textit{BiD} under  $\left(\bm{p}^{\text{W}\star},\Phi_{-W}^{\text{DA}\star}\right)$,  meaning that we always have $S^{\text{BiD}}\geqslant S^{\text{BiD-q}}$. \qed


\bibliographystyle{apalike} 
\bibliography{storage.bib}

\end{document}